\documentclass[conference]{IEEEtran}
\IEEEoverridecommandlockouts

\usepackage{cite}
\usepackage{amsmath,amssymb,amsfonts}
\usepackage{algorithmic}
\usepackage{graphicx}
\usepackage{subcaption}
\usepackage{textcomp}
\usepackage{xcolor}
\def\BibTeX{{\rm B\kern-.05em{\sc i\kern-.025em b}\kern-.08em
    T\kern-.1667em\lower.7ex\hbox{E}\kern-.125emX}}

\makeatletter
\newcommand{\linebreakand}{%
  \end{@IEEEauthorhalign}
  \hfill\mbox{}\par
  \mbox{}\hfill\begin{@IEEEauthorhalign}
}
\makeatother

\begin{document}

\title{Exploration of Evolving Quantum Key Distribution Network Architecture Using Model-Based Systems Engineering
\thanks{This work was partly supported by the Innovate UK [grant number 10102791]}
}
\author{
\IEEEauthorblockN{Hayato Ishida\IEEEauthorrefmark{1}\IEEEauthorrefmark{3}, Amal Elsokary\IEEEauthorrefmark{1}\IEEEauthorrefmark{3},
Maria Aslam\IEEEauthorrefmark{1}, Catherine White\IEEEauthorrefmark{2}\\,
Michael J. de C. Henshaw\IEEEauthorrefmark{1}, \IEEEmembership{Member}, IEEE, Siyuan Ji\IEEEauthorrefmark{1}, \IEEEmembership{Member}, IEEE}
\IEEEauthorblockA{\IEEEauthorrefmark{1}Loughborough Quantum Systems Research Group, Loughborough University,\\ Loughborough, UK, \{h.ishida, a.elsokary, m.aslam, m.j.d.henshaw, s.ji\}@lboro.ac.uk \\ \IEEEauthorrefmark{2}BT Research, Ipswich, UK, catherine.white@bt.com \\ \IEEEauthorrefmark{3}Both authors contributed equally}
}












\maketitle

\begin{abstract}
Realisation of significant advances in capabilities of sensors, computing, timing, and communication enabled by quantum technologies is dependent on engineering highly complex systems that integrate quantum devices into existing classical infrastructure. 
A systems engineering approach is considered to address the growing need for quantum-secure telecommunications that overcome the threat to encryption caused by maturing quantum computation. 
This work explores a range of existing and future quantum communication networks, specifically quantum key distribution network proposals, to model and demonstrate the evolution of quantum key distribution network architectures. 
Leveraging Orthogonal Variability Modelling and Systems Modelling Language as candidate modelling languages, the study creates traceable artefacts to promote modular architectures that are reusable for future studies. 
We propose a variability-driven framework for managing fast-evolving network architectures with respect to increasing stakeholder expectations. 
The result contributes to the systematic development of viable quantum key distribution networks and supports the investigation of similar integration challenges relevant to the broader context of quantum systems engineering.
\end{abstract}


\section{Introduction}
The UK has invested over £1 billion in quantum technologies since 2012 and is committed to a further £2.5 billion over the next ten years \cite{wilkes2025quantum}. 
Quantum technologies offer massive capability advances and investment opportunities if significant challenges can be overcome. 
A major challenge is to move from laboratory demonstration to practical devices and systems; this will be a focus for future investment, and the role of Systems Engineers in managing the concomitant complexity cannot be understated. 
In this paper, we present an approach through which Model-Based Systems Engineering (MBSE) can be applied to a contemporary quantum application, to clarify the function and interfaces of quantum mechanical elements in the system in a way that provides an accessible high-level view even to the non-expert.

Quantum Key Distribution (QKD) network systems use quantum principles, such as superposition and entanglement, to enable secure information transfer across distributed nodes. 
A main use case for QKD network is to provide a communication system that is secure against quantum computers, for which QKD systems in principle provide an unconditionally secure mitigation, while future quantum communication networks (QCN) are anticipated to enable the transfer of quantum information between quantum computers. 
Leveraging a unified communication infrastructure with quantum and classical channels, which are integrated to provide a combined function, enables a wide range of use cases, such as in QKD protocols, which require both. 
Implementation may even involve physical commonalities such as co-propagation of quantum and classical cryptography on a communication medium, potentially reducing costs but at the risk of crosstalk from the classical to the quantum channel \cite{tan2024simultaneous}. 

It is acknowledged that the integration of quantum technologies into existing communication infrastructure presents formidable challenges \cite{orieux2016recent}, partly attributed to the technological complexity involved in quantum subsystems, such as their control and the management of their interfaces with the classical environment in which they operate. 
This led to the proposal of taking tailored systems engineering approaches to manage such complexity \cite{henshaw2017challenges}, which later evolved into a distinct, cross-disciplinary field, referred to as quantum systems engineering (QSE) \cite{everitt2016quantum}. 
However, this discipline has not gained sufficient momentum until recently, with recognition of the need to take a structured approach to solving identified engineering challenges \cite{IET2024QuantumSE}. 
A particular research direction in QSE is the investigation of how MBSE could support the management of quantum systems complexity, in the same way it has been successfully applied to other complex systems (e.g. in the defence and aerospace context). 
Given that quantum physicists and engineers present their QKD network architecture proposals with specialised language and drawings \cite{chen2009field, peev2009secoqc, sasaki2011field, huang2025fully, dianati2008architecture} that are unlikely to be accessible to a wide range of stakeholders, particularly all systems engineers, this naturally motivated the idea of applying a standardised and universal modelling language, such as the Systems Modelling Language (SysML), to capture such proposals. 

A few significant challenges have been encountered in such an investigation. Firstly, the effort required to model a QKD network architecture proposal can be substantial, because the amount of knowledge required goes beyond a single discipline, i.e., quantum physics, network architecture, and MBSE, to state the minimum. 
Secondly, proposals clearly show repeated patterns. However, without a holistic and tailored approach to identify and model these patterns specific to quantum systems, reusing model elements from architecture to architecture requires heavy manual efforts. 
Lastly, the majority of these architectures are not designed against clear stakeholder needs. In fact, the novelties are often in the detailed design, rather than at the architectural level. This makes a standard top-down approach to systems architecting inappropriate. The aim of this work is to address this challenge and share our findings.

The contribution of this work can be summarised as a validated framework that can be specifically used to effectively model various QKD network architectures to promote reusability of modular architecture, agility in composing new architecture against stakeholder needs, and effectiveness in communicating the architecture to different stakeholders. 

This paper is structured as follows. Section \ref{sec:background} presents the essential background of this work. The framework is then presented in Section \ref{sec:framework}, followed by a validation of the framework in Section \ref{sec:validation}, which also includes a discussion on the practicality, adaptability and scalability of the approach presented by the framework. 
The final two sections discuss related work and offer a conclusion, respectively.

\section{Background}

This section provides essential background knowledge for this work. 
\label{sec:background}
\begin{figure}[htbp]
\centerline{\includegraphics[width=7.5cm]{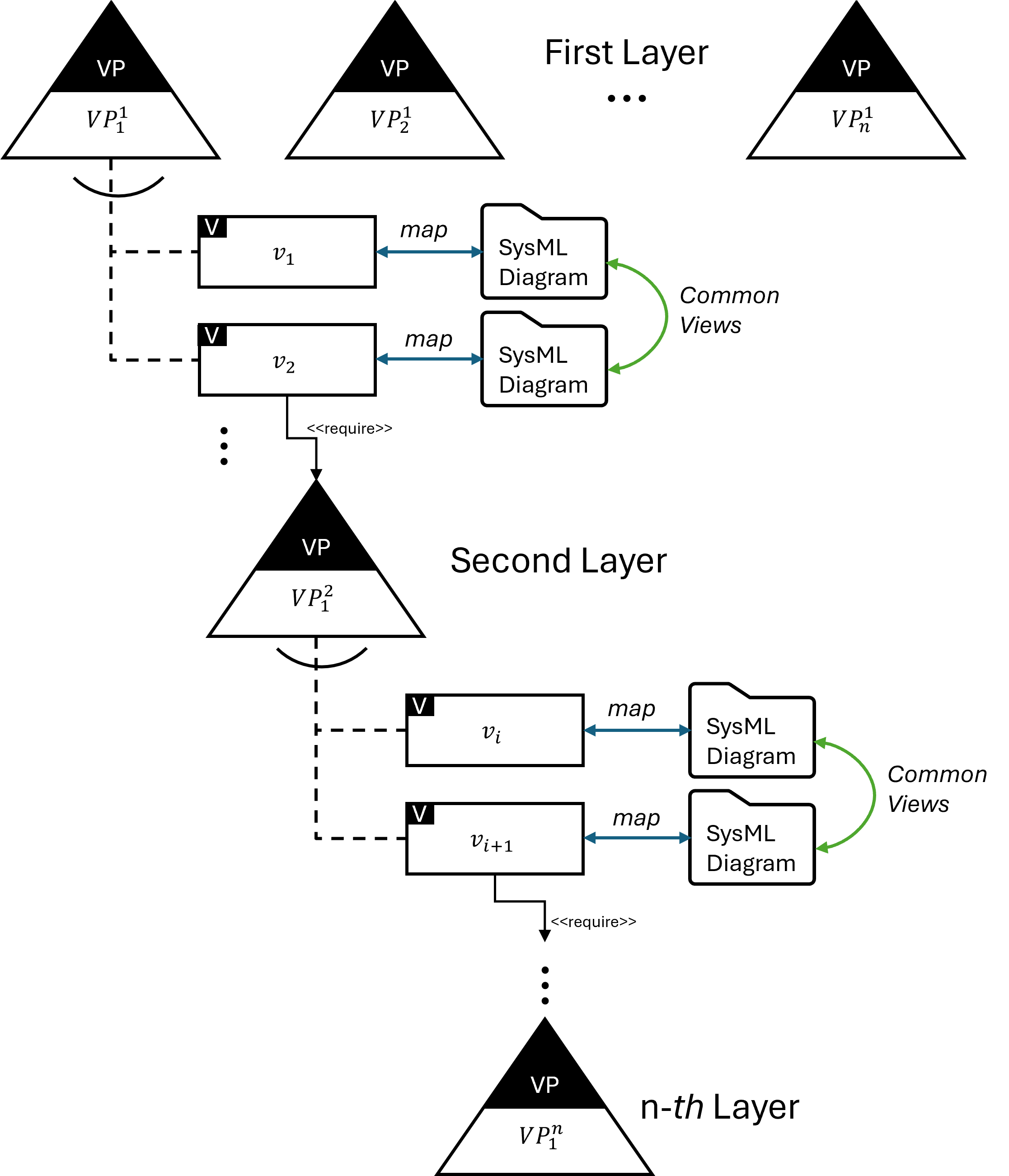}}
\caption{A hierarchical OVM+SysML model structure without quantum-specific contents.}
\label{fig:framework_structure}
\end{figure}

\subsection{Quantum key distribution}
QKD, first introduced by Bennett and Brassard in 1984 \cite{BennettBrassard1984}, 
known as the BB84 protocol, has been one of the most active research fields in quantum information.
A succession of protocols has since been developed, notably Ekert’s entanglement‑based scheme (E91) \cite{ekert1991quantum}, and the measurement device-independent (MDI) QKD \cite{Lo2012}, each addressing distinct security and implementation challenges of previous protocols.

For any QKD protocol, it follows the following steps:
\begin{enumerate}
    \item \textbf{Quantum transmission}: the legitimate parties (Alice and Bob) exchange quantum states.
    \item \textbf{Sifting}: public comparison of classical metadata (e.g., encoding and measurement bases) yields a correlated \emph{raw key}.
    \item \textbf{Post‑processing}: classical procedures such as error correction and privacy amplification produce an identical, \emph{secret key}.
\end{enumerate}
Although the overarching framework is universal, individual protocols vary in their state‑preparation procedures, measurement methodologies, and the reconciliation techniques applied during post‑processing.
QKD utilises the laws of quantum mechanics to carry out private key distribution between two users.
Because security relies on fundamental quantum laws, principally the no-cloning theorem and contextuality, QKD offers information-theoretically secure cryptography.
By contrast, RSA, which is one of the most used classical cryptographic protocols, is computationally secure, meaning that it is dependent on the assumption of computational hardness for security.
Any eavesdropping attempt necessarily induces detectable disturbances, enabling the communicating parties to detect the presence of an eavesdropper and abort the protocol when the channel is compromised.

\subsection{Model-based systems engineering}
Model‑based systems engineering (MBSE) expresses engineering data through formal models rather than narrative documents \cite{iso15288:2023}.
This approach enhances traceability, supports explicit management of dependencies and interfaces, and mitigates miscommunication within interdisciplinary teams—a capability of particular importance to quantum technologies. 
A further benefit of MBSE is its modularity: each subsystem is represented as an independent module, which can be reused or evolved with minimal impact on its neighbours.
Such a "divide-and-conquer" approach allows engineers to grasp individual modules before addressing system‑level interactions, thereby reducing overall complexity.

The Systems Modelling Language (SysML) is the prevailing standard for MBSE. All models in this study were developed in SysML 1.7; adoption of the recently released SysML v2 is expected to grow as modelling tools mature.

\subsection{Orthogonal variability modelling}
Orthogonal variability modelling (OVM), widely employed in product‑line engineering, captures variation points, associated variants and their interrelationships \cite{hause2019model}. 
A variation point denotes a system feature for which several implementation options exist.
OVM is therefore well suited to QKD networks, where multiple protocol choices and corresponding architectural implications must be represented. 
Moreover, the method can portray prospective configurations, thereby supporting forward‑looking analysis of alternative system evolutions.

\section{The Framework}
\label{sec:framework}
\subsection{Model Development Process}
\label{sec:process}

\begin{figure}[b]
\centerline{\includegraphics[width=7.5cm]{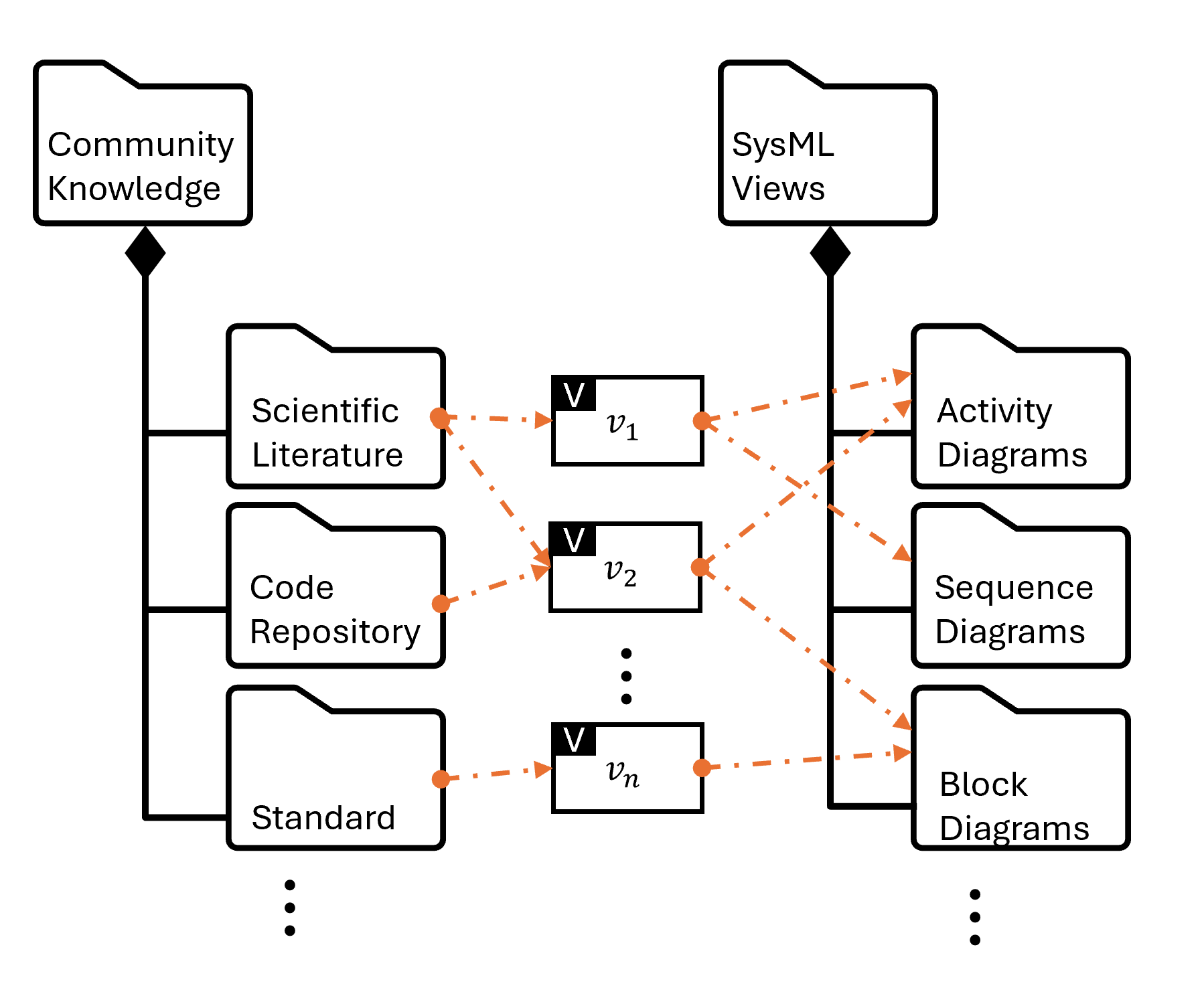}}
\caption{Illustrative model development process (Steps 1 and 2 only) using quantum-specific knowledge.}
\label{fig:framework_process}
\end{figure}

\begin{figure}[t!]
\centerline{\includegraphics[width=8.5cm]{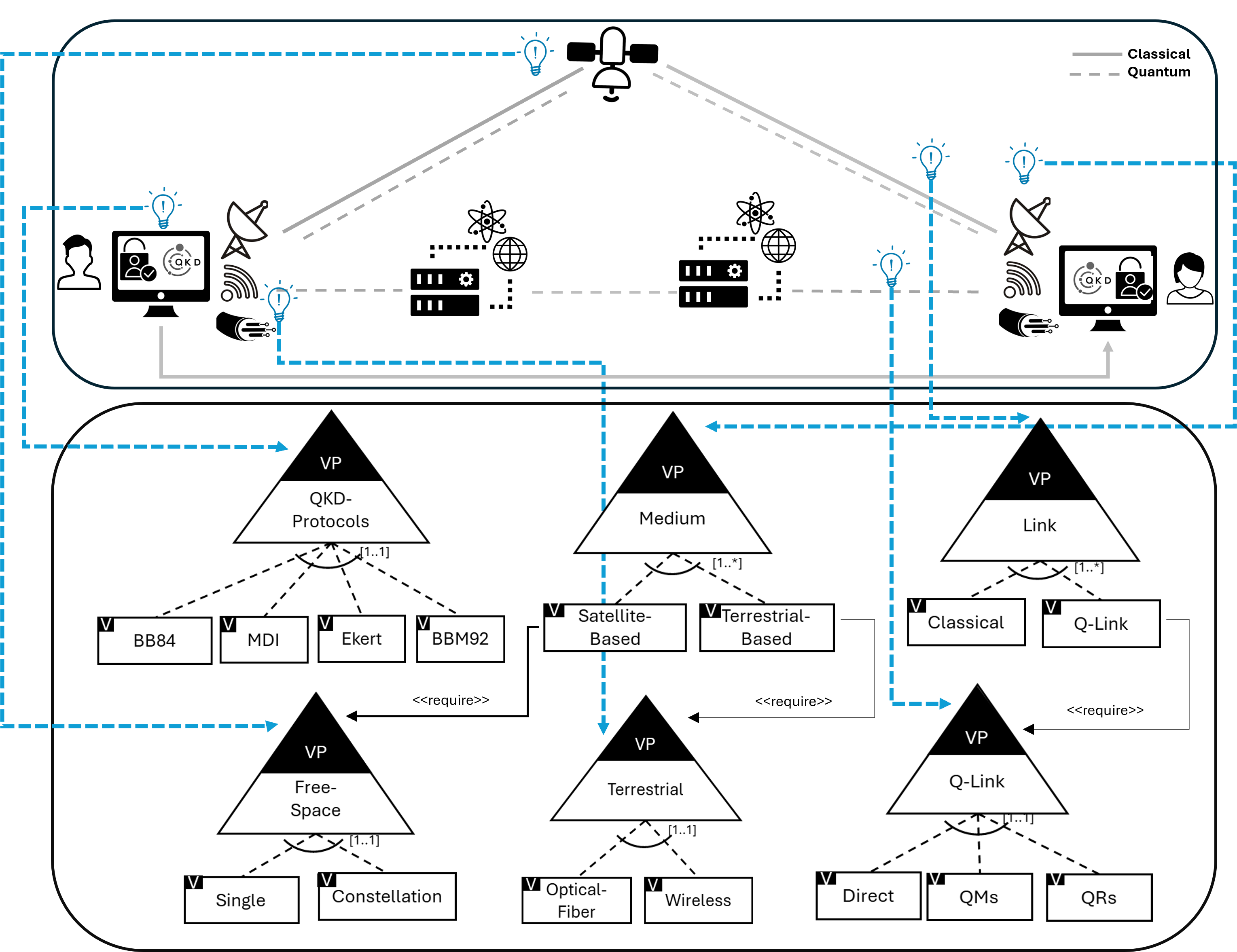}}
\caption{OVM backbone derived from domain-specific knowledge}
\label{fig: OVM_backbone}
\end{figure}

\begin{figure}[h!]
\centering
    \begin{minipage}[t]{0.3\textwidth}
        \centering
        \includegraphics[width=\linewidth]{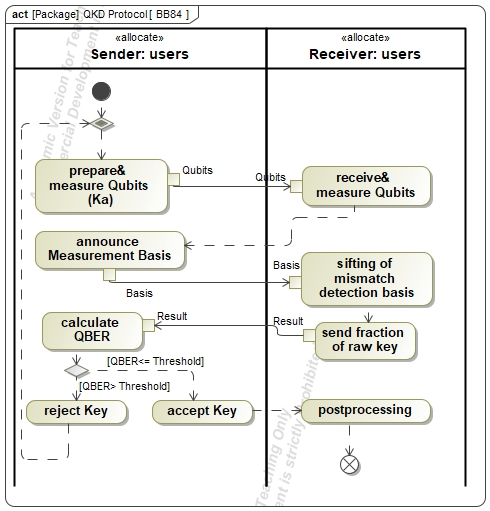}
        \caption*{(a) BB84 QKD}
    \end{minipage}
    
    \begin{minipage}[t]{0.3\textwidth}
        \centering
        \includegraphics[width=\linewidth]{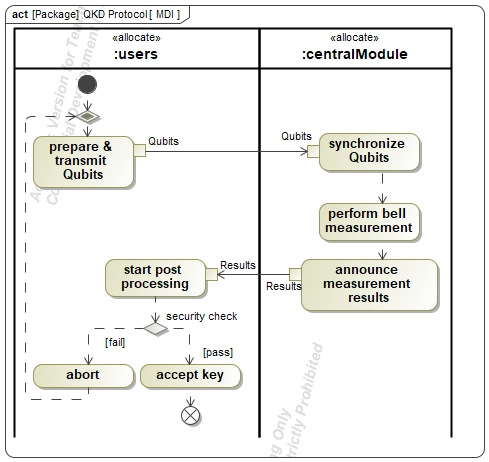}
        \caption*{(b) MDI-QKD}
    \end{minipage}

    \begin{minipage}[t]{0.4\textwidth}
        \centering
        \includegraphics[width=\linewidth]{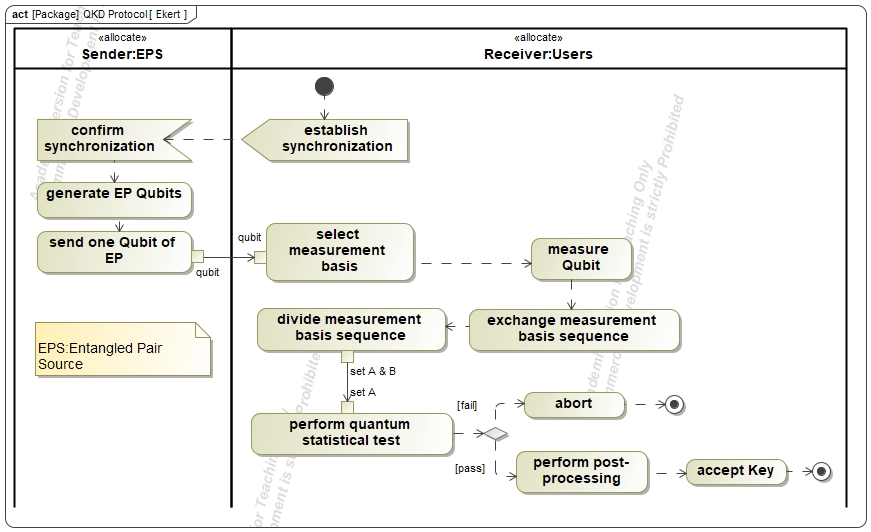}
        \caption*{(c) Ekert-QKD (E91)}
    \end{minipage}
    
\caption{QKD protocol processes.}
\label{fig:qkd_protocols}
\end{figure}

To explore the evolving QKD network architecture and organise the complexities observed in the architectural designs, inspired by reviewed works, specifically Model-Based Product Line Engineering (MBPLE) by \cite{li_model-based_2015} and the variability-realisation mechanism by \cite{schafer_variability_2021}, we present a model-based QKD network architecture exploration framework in this section. To avoid confusion, we highlight that this framework is not a typical architectural framework (e.g., the Unified Architectural Framework), but outlines an approach for how a range of known QKD network architectural entities, proposed or realised, could be modelled by combining OVM and SysML, to support the investigation and future development of a QKD network system. 

\subsection{Model Structure}
The first part of the framework is a non-quantum specific model structure developed and presented in Fig. \ref{fig:framework_structure}. As shown, within a standard multi-layered hierarchical OVM \cite{hause2019model}, for each of the variants in the OVM, a SysML-based diagram is developed to provide a model-based specification of that variant. This creates a one-to-one mapping between the OVM and the SysML-based model. Then, for a group of variants under a variation point, a specific diagram type, e.g., Activity Diagram, that is most appropriately aligned with the concept of that variation point, will be utilised to model the range of variants. This allows the variants to share a 'common' view, with similar structures, such that stakeholders could explore the differences between these variants without the need to investigate other variation points. 
We highlight that this framework presents an approach opposite to the one proposed by Li \cite{li_model-based_2015}, where variation points and variants are derived from SysML-based system models. In our framework, the SysML-based models are developed after the variation points and variants are defined. This approach is supported by the following fact: Unlike in the fields of software, automotive and aerospace, where system models are often abundant, due to the industrial uptake of model-driven development and MBSE, as reviewed in the related work section, there is very limited work in model-based QSE. 



    

    



The second part of the framework is a process, shown in Fig. \ref{fig:framework_process}, that enables the development of a baseline model consisting of the OVM backbone and the set of SysML diagrams that map to the OVM variants. This part is quantum-specific. 

Following the construct of the framework model structure, one would naturally start the model development by defining the OVM backbone following a top-down approach, where one would identify and specify the first layer variation points and the associated variants. Subsequently, one would follow the model structure in Fig. \ref{fig:framework_structure} downward to complete the hierarchy until a sufficient level of engineering detail is reached. However, an initial investigation shows that this intuitive top-down approach did not work well. This is primarily because the evolution of quantum technologies is found to be driven by breakthroughs at the component level rather than at the system level, e.g., quantum memory device \cite{moiseev2025optical}. For this reason, we propose an iterative, experimental, bottom-up process for developing the OVM+SysML model. 

\textbf{Step 1 - Knowledge Gathering}: This step is concerned with gathering initial domain knowledge for the QKD network with sufficient details to enable the next step, but does not need to be exhaustive. For instance, this would include the latest experimental results on the realisation of quantum behaviours that promise commercial viability, validated GitHub repositories that provide digital twins of quantum devices, and community-accepted standards that benchmark quantum technologies. 

\textbf{Step 2 - Model Development}: Using the gathered knowledge as the basis, we then derive an initial list of non-grouped 'variants' that represent key quantum features, e.g., quantum teleportation. Then, for each of the variants, we test using different SysML Diagrams to capture the core concepts of that variant. After having an initial set of views (SysML diagrams) developed, we then analyse the variants and the view holistically to:
\begin{enumerate}
    \item identify structural and behavioural commonalities between variants.
    \item group variants accordingly to define variation points that generalise the concepts.
    \item normalise the views for each variation point such that a single SysML diagram is used consistently across the group of variants. 
\end{enumerate}

\textbf{Step 3 - Iteration}: Acknowledge that there could be many ways of organising the knowledge into an OVM backbone structure, which would lead to different ways in normalising the SysML diagrams against the variants, Step 3 is therefore about experimenting with different OVM backbone structures with potentially added knowledge that capturing most recent breakthroughs in quantum technology development. This step is dominated by trial-and-error, featuring an iterative development process. While it can be difficult to optimise the OVM backbone structure, we recommend that principles for the experimentation be defined and used in driving this step. For instance, the grouping should balance between a structure that is sensible to quantum physicists and is intuitive for a system engineer to model in SysML. 

\textbf{Step 4 - Configuration}: Once the OVM+SysML model is completed, configurations can be composed against stakeholder needs. Specifically, for any given stakeholder's needs, instead of developing a QKD network architecture from scratch, the systems engineers could work with quantum engineers to quickly compose a meaningful configuration through the selection of a set of variants. Then, the initial architectural description would be the composition of the set of corresponding SysML diagrams. We highlight that this composed architectural description will not fully represent a realistic QKD network. This is because the OVM+SysML model structure does not address the interactions between variants. As such, this step also requires the necessary modification of individual SysML diagrams to ensure consistency between views and the creation of a configuration-specific interface definition.

\section{Validation}
\label{sec:validation}

The framework is validated in this section to show its applicability and scalability. Firstly, a baseline OVM+SysML model is developed guided by the framework proposed in the previous section. This is then followed by a case study, which demonstrates how an initial architecture can be quickly composed to address a high-level stakeholder need. The section is concluded with a discussion of the benefits and limitations of the proposed framework. 

\subsection{The OVM Backbone and SysML Views}


Firstly, the initial list of variants is defined based on a broad review of relevant works, such as Refs \cite{granelli2022novel, huang2025fully, sasaki2011field, peev2009secoqc}. These variants are shown in Fig.~\ref{fig: OVM_backbone}. The groupings of the variants define the variation points (VP) and their hierarchical structure, also shown in the diagram. As an example, different types of communication media are grouped into a 'Medium' VP, and different satellite arrangements are grouped into a 'Free-Space' VP. The two VPs are hierarchically linked through the 'Satellite-based' medium variant, indicating the need for further specification of satellite arrangement if the 'Satellite-based' medium variant is selected. To make sense of the VPs for the readers, a schematic network diagram is drawn above the OVM backbone, showing what the VPs are trying to abstract. Before arriving at this final version, a range of intermediate OVM backbone structures were derived but discarded based on stakeholder feedback, as guided by Step 3 of the process. It is worth highlighting that not all VPs and variants are quantum-specific. This makes it challenging to decide on the final OVM backbone structure in terms of balancing different stakeholder views. 

The other concurrent activity in Step 2 concerns developing SysML views for the variants. Again, instead of showing the intermediate views developed, we only show the final, normalised views that align with the OVM backbone. 

In Fig.~\ref{fig:qkd_protocols}, QKD protocols, BB84 (Fig.~\ref{fig:qkd_protocols}(a)), MDI-QKD (Fig.~\ref{fig:qkd_protocols}(b)) and Ekert (Fig.~\ref{fig:qkd_protocols}(c)) are modelled using SysML activity diagrams. The sequence diagram was also tested as a possible candidate in previous iterations (Step 3 of the process). However, the activity diagram is chosen as the final type because the interaction mechanism of complicated protocols (e.g., Ekert) can lead to a lengthy sequence diagram that makes it challenging to normalise with protocols that are simpler (e.g., BB84). The contents of the activity diagram are confined to operational processes, rather than showing details of the behaviours.

\begin{figure}[htbp]
\centering
    \begin{minipage}[b]{0.25\textwidth}
        \centering
        \includegraphics[width=\linewidth]{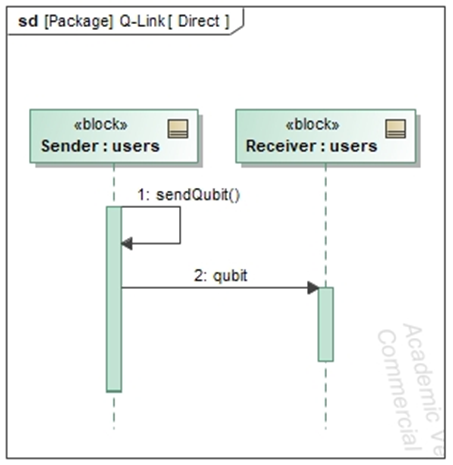}
        \caption*{(a) Direct Q-Link}
    \end{minipage}
    \hfill
    \begin{minipage}[b]{0.4\textwidth}
        \centering
        \includegraphics[width=\linewidth]{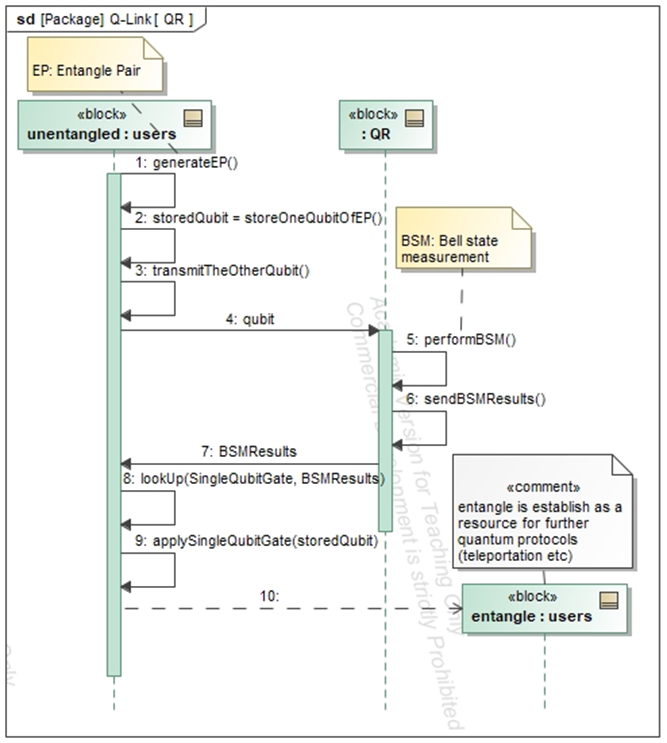}
        \caption*{(b) Q-Link with Entanglement Swapping}
    \end{minipage}
\caption{Quantum Link Variability}
\label{fig:q_link}
\end{figure}

Fig.~\ref{fig:q_link} illustrates the Q-link (quantum link) VP and highlights the different communication paths, where Fig.~\ref{fig:q_link} (a) depicts the direct quantum channel between two users, a sender and a receiver. 
However, direct Q-link is only practical over short distances. The other sequence diagram models how long-distance communication can be established by using quantum repeaters \cite{azuma2023quantum, liorni2021quantum},  which is a more advanced quantum communication technology yet to be fully realised. The sequence diagram is regarded as the most appropriate diagram type to be used for this VP because the link mechanism is meant to describe how information is exchanged, and the sequence diagram, as an interaction diagram, is used exactly for modelling information exchange. 

\begin{figure}[h!]
\begin{minipage}[b]{0.23\textwidth}
    \includegraphics[width=\linewidth]{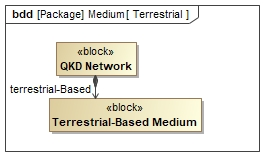}
    \caption*{(a) Terrestrial-based medium}
\end{minipage}
\hfill
\begin{minipage}[b]{0.23\textwidth}
    \includegraphics[width=\linewidth]{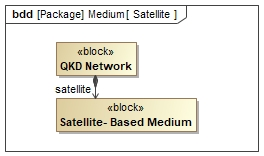}
    \caption*{(b) Satellite-based medium.}
\end{minipage}
\caption{QKD network medium.}
\label{fig:qkd_medium}
\end{figure}

Fig.~\ref{fig:qkd_medium} shows the modelling of the medium VP, utilising the SysML block definition diagram to capture the hierarchical structure in a comparative manner. The choice of block definition diagram is straightforward, as it is the most convenient diagram type to capture system hierarchy. 


\subsection{Case Study: Long Range Communication}
To demonstrate the effectiveness of composing an initial architectural description from the OVM+SysML model, we carried out the following case study. 

A synthetic stakeholders’ use case is formulated as follows: A secure network is needed for long-range communication. The network must also be resilient; thus, it shall have redundancies that would ensure the network continues operating securely if one of its channels is attacked. QKD is the preferred technology to test feasibility. 

\begin{figure}[b]
    \begin{minipage}[b]{0.22\textwidth}
        \includegraphics[width=\linewidth]{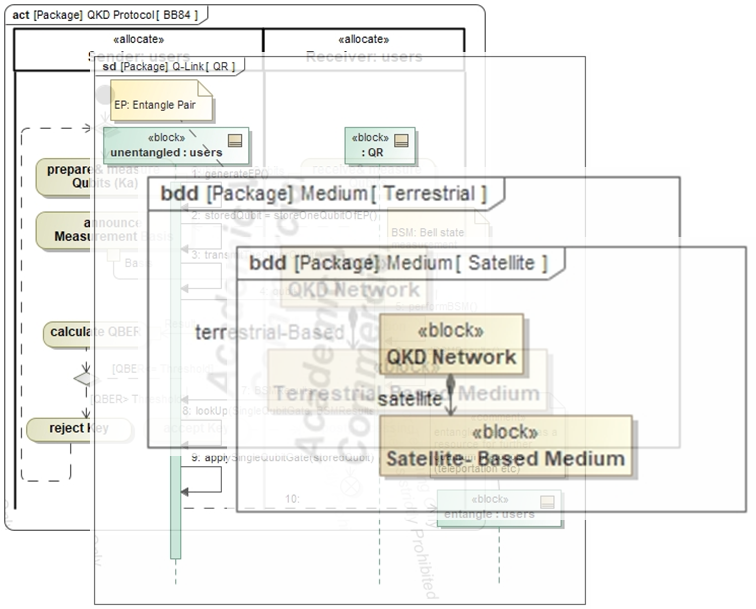}
        \caption*{(a) Composed SysML views}
        \label{fig:configuration_initial}
    \end{minipage}
    \begin{minipage}[b]{0.23\textwidth}
        \includegraphics[width=\linewidth]{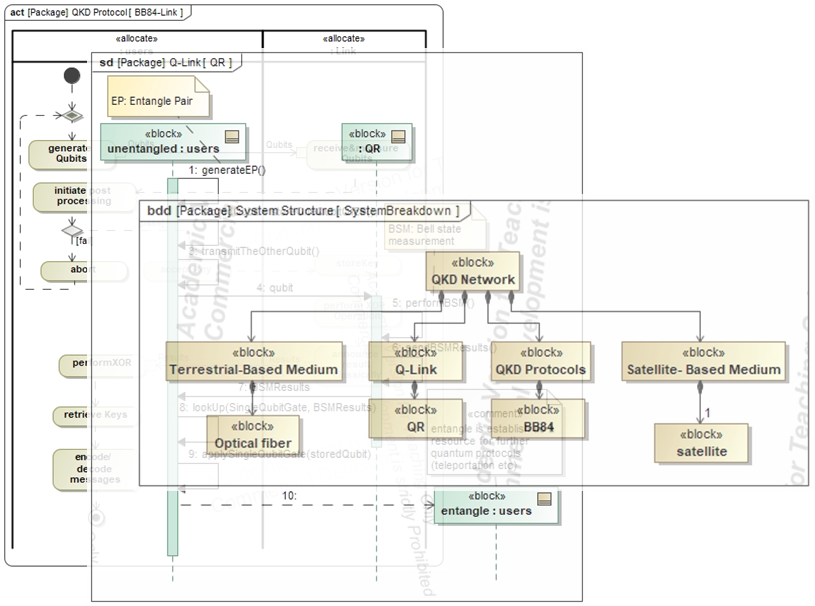}
        \caption*{(b) Modified SysML views}
        \label{fig:configuration_modified}
    \end{minipage}
\caption{Initial QKD network architecture description.}
\label{fig:qkd_network_architecture}
\end{figure}

\begin{figure}[t]
\centerline{\includegraphics[width=8cm]{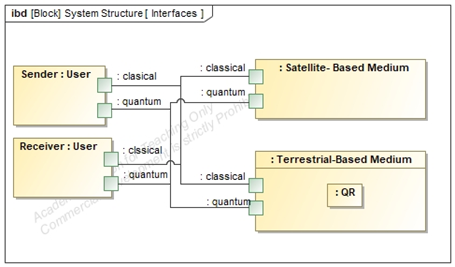}}
\caption{Interface definition for the composed configuration.}
\label{fig:interface_definition}
\end{figure}

In discussion with quantum systems engineers, these stakeholder needs have led to the selection of the following variants:
\begin{itemize}
    \item For the Medium VP, both satellite and terrestrial variants are selected to ensure the network will still be available in case one of the channels fails or is being attacked:
    \begin{itemize}
        \item For the required Free-space VP, the ‘Single’ variant is selected, i.e., one satellite
        \item For the required Terrestrial VP, the optical fibre variant is selected
    \end{itemize}
    \item For the QKD Protocol VP, the BB84 protocol is selected
    \item For the Q-Link VP, QR is selected as the direct quantum link will not achieve system fidelity
\end{itemize}
Fig.~\ref{fig:qkd_network_architecture} and Fig.~\ref{fig:interface_definition} illustrate the results obtained from carrying out Step 4 of the process. The views associated with the selected variants are composed, as depicted in Fig.~\ref{fig:qkd_network_architecture}(a). These views then modified into a set of new views to enhance consistency, as in Fig.~\ref{fig:qkd_network_architecture}(b), serving the initial behavioural and structural architectural description for the configuration. 
The revised model finally enables the interface definition for this particular QKD network architecture. This interface definition is captured using an internal block definition diagram, as shown in Fig.~\ref{fig:interface_definition}.

\subsection{Discussion}
\label{sec:validation_discussion}
\textit{Practicality}- The successful development of the OVM+SysML model guided by the framework, together with the successful completion of the case study, proves the practicality of the framework in the exploration of QKD network architecture. However, it is also clear that both the model and the case study are simple validations for the purpose of demonstration. A more comprehensive validation would require a number of case studies with different stakeholder needs that lead to different configurations. Such a validation is necessary before the framework to be deployed for industrial-level studies. 

One could argue that to enhance the practicality of the framework, capabilities such as architecture evaluation against the stakeholder needs, though out of scope for this work, could be desirable for a future version. 

\textit{Adaptability}- Although the framework is developed specifically for the exploration of QKD network architecture, it is observed that the framework could be adapted to other types of systems that utilise different quantum technologies, and could be potentially made domain-independent. This is because the model structure proposed in the framework, as shown in Fig.~\ref{fig:framework_structure}, is non-quantum specific, meaning that one could, in principle, interpret any knowledge into it, following a structure-preserving paradigm discussed in \cite{ji2022structure}. 

However, when reflecting on the framework process, we highlight that the framework is much less useful to domains in which system architecture is significantly more mature, e.g., aircraft that use jet engines. This is because the primary capability of this framework is to enable the exploration of unrealised system architecture through a bottom-up approach. 

\textit{Scalability}- A typical concern for any model-based approach for systems engineering is its scalability against increasing system complexity. This is because the number of model elements (including relationships) increases quickly as the system size grows. Methods that rely on manual processes are less likely to be scalable. We argue that our framework offers scalability in the sense that the OVM backbone can be extended by adding more VPs or variants at any time, with respect to new breakthroughs. This is a feature inherited from using OVM to construct the product line model. The challenge in scaling the OVM potentially lies with the need for a completely new backbone structure, perhaps due to a disruptive technology breakthrough that leads to a void of legacy variants. 

The part that is not as scalable is the configuration process. As the number of VPs and variants grows, the manual intervention required to refine the views for consistency could become infeasible. To address this challenge, as a potential future work, an automation mechanism to enable synchronisation of views \cite{ji2023requirements} could be developed. The automation mechanism could consider extending the OVM backbone structure to include dependencies between variants, such that the growth of the configuration space size is manageable \cite{li2019product}. 

\section{Relevant Works}

\subsection{Quantum Systems Engineering}
The need for a systems engineering approach in the development of quantum systems or technologies has been discussed previously
\cite{everitt2016quantum, henshaw2017challenges, bjergstrom2020quantum}, with those studies principally addressing the challenges that QSE must resolve.
QSE differs from a direct application of systems engineering to quantum systems. 
It is not simply a matter of scaling familiar methods to a more complex system or technology: quantum technologies operate under entirely different physical laws, where superposition, entanglement, and quantum measurement replace the deterministic behaviour assumed in classical systems.
Various challenges for quantum systems engineering arise due to the unique characteristics of the quantum domain, particularly entanglement.
Everitt et al. \cite{everitt2016quantum} identify two principal challenges: the interface problem and the modelling problem. 
The interface problem arises from entanglement, which prevents treating subsystems as isolated, obstructing the definition of clear boundaries, interfaces, and requirements.
Quantum systems inhabit Hilbert spaces, and tensor‑product composition causes the system’s degrees of freedom to increase exponentially, complicating modelling and simulation.
The modelling problem also arises from the exponential growth of the simulating complexity of the quantum system \cite{Feynman1982, Troyer2005}. 
A third obstacle, environment coupling, introduces an additional challenge: unavoidable interactions with the surroundings cause decoherence and blur the very system boundaries that classical engineering relies on for modular design \cite{bjergstrom2020quantum}.
There are additional challenges concerning testing and scalability.
Furthermore, the work by Henshaw et al. \cite{henshaw2017challenges} maps these challenges and issues of quantum systems engineering against technical processes defined by the ISO/IEC/IEEE 15288:2015 \cite{iso15288:2015}. 
Their mapping shows specific shortfalls in requirements capture, architectural definition, integration, verification and validation when those processes confront quantum phenomena.
Thus, the challenges in terminology are familiar to the systems‑engineering community.

\subsection{UML and SysML for Quantum Systems}
Several previous studies have addressed the challenges of modelling quantum systems.
Although quantum software engineering differs from quantum systems engineering (QSE), it shares foundational concepts such as the Unified Modelling Language (UML) and the Systems Modelling Language (SysML). 
Silverman and Jiron \cite{Silverman2023} present the use of MBSE and SysML for quantum systems with a focus on quantum computing and architecture.
Research applying UML to quantum software is more extensive than research in QSE. 
Prior studies have used UML to model quantum software or algorithms \cite{ perez2022design, guo2025quanuml}. 
These studies concentrate on quantum circuit-level models, which may not scale to the thousands of qubits anticipated in future quantum computers. 
Work by P{\'e}rez-Castillo and Piattini \cite{perez2022design} extends UML for quantum software; however, this extension cannot be applied directly to broader quantum systems that include hardware elements.

Although previous studies have aimed to model quantum systems, prior efforts have not fully resolved the challenges identified by Everitt et al \cite{everitt2016quantum}. 
This outcome is unsurprising because the cited studies address quantum software rather than complete quantum systems.
Consequently, a substantial research gap persists.
This paper addresses some of the SysML modelling challenges for quantum systems; however, precise SysML representation of uniquely quantum features (e.g., entanglement) remains a topic for future work.

\subsection{Model-Based Product Line Engineering}
QKD network systems are complex because they integrate both classical and quantum components, and the variety of QKD protocols and their architectures is continually expanding. 
To manage this complexity and variability, this study investigates MBPLE, the fusion of MBSE’s system‑wide traceability with PLE’s variability management.
MBSE provides engineers with a clear, consistent model of the whole system, but its built‑in support for variation points is limited; consequently, engineers frequently duplicate models when many product variants are needed \cite{li_model-based_2015}.
Conversely, PLE captures variability via feature or option models but lacks explicit traceability from each selection to behavioural, structural, and verification artefacts \cite{dumitrescu_capturing_2014}.
Combining the two approaches yields a single traceable system model while enabling safe selection and reuse of variants \cite{schafer_variability_2021}.

Among the variability‐realisation mechanisms catalogued by Schäfer et al.\cite{schafer_variability_2021}, three mechanisms are particularly well suited to a QKD‑network product line: 
\begin{itemize}
    \item Module replacement enables engineers to substitute protocol‑specific devices (e.g., decoy‑state sources or entangled‑photon detectors) without modifying the baseline classical control model. 

    \item Conditional compilation retains alternative security primitives within a single over‑specified SysML model and selects the appropriate primitive at build time.

    \item Polymorphism associates different key‑management algorithms with a common interface, preserving higher‑level orchestration.
\end{itemize}

Complementing these design‑time tactics, Li et al.’s requirement‑level variability modelling propagates feature selections from stakeholder needs through functional, logical and physical views, preserving end‑to‑end traceability—an essential property for security‑accredited QKD deployments \cite{li_model-based_2015}.
Variation points are specified once in the requirements view and then propagated into block, behavioural and parametric diagrams, preserving traceability without additional diagrammatic clutter \cite{li_model-based_2015}.

For a QKD network, as shown in previous sections, variant choices such as protocol family and quantum-link may be specified once and automatically propagated through functional, logical, and physical SysML layers, preserving consistency and traceability across every network variant \cite{li_model-based_2015}.

\section{Conclusion and Future Works}

This work presented a model-based framework for exploring fast-evolving QKD network architectures while addressing stakeholder expectations. The approach enables stakeholders to investigate the impact of adopting quantum technology in their systems without committing to fully developing requirements and architecture. By leveraging OVM, the framework enables rapid creation and assessment of configurations.

Currently, QKD network technologies remain in early development, with few practical implementations. As quantum communication networks mature, this framework will serve as a valuable tool for guiding their integration into classical infrastructure. To fully enable this envisaged capability, a natural future course is to expand the current OVM+SysML model to include a wider range of architectural entities, such that more complex configurations can be efficiently composed. The models can be accessed via the Loughborough Quantum Systems Engineering GitHub repository \cite{Ishida2025QCNModels}. 

In addition to the future work highlighted in the discussion in Section \ref{sec:validation_discussion}, we also highlight an additional opportunity:  investigation of utilising SysMLv2 capabilities to enhance the framework, including, for instance, the development of a comprehensive modelling profile or library that supports precise modelling of quantum devices and their behaviours.

\section*{Acknowledgement}
C. White, M. Henshaw and S. Ji acknowledge support by QAssure, an Innovate UK-funded project, part of the Quantum Challenge within the UK National Quantum Technologies Programme.

\section*{CRediT Authorship Contribution Statement}
\noindent
\textbf{Hayato Ishida:} Investigation, Methodology, Visualisation, Writing - Original draft. 
\textbf{Amal Elsokary:} Investigation, Methodology, Visualisation, Writing - Original draft. 
\textbf{Maria Aslam:} Writing - Review and Editing.
\textbf{Catherine White:} Validation, Writing - Review and Editing. 
\textbf{Michael J. de C. Henshaw:} Supervision, Writing - Review and Editing. 
\textbf{Siyuan Ji:} Conceptualisation, Supervision, Writing - Original draft, Writing - Review and Editing.

\bibliography{IEEEabrv,references}

\begin{thebibliography}{10}
\providecommand{\url}[1]{#1}
\csname url@samestyle\endcsname
\providecommand{\newblock}{\relax}
\providecommand{\bibinfo}[2]{#2}
\providecommand{\BIBentrySTDinterwordspacing}{\spaceskip=0pt\relax}
\providecommand{\BIBentryALTinterwordstretchfactor}{4}
\providecommand{\BIBentryALTinterwordspacing}{\spaceskip=\fontdimen2\font plus
\BIBentryALTinterwordstretchfactor\fontdimen3\font minus \fontdimen4\font\relax}
\providecommand{\BIBforeignlanguage}[2]{{%
\expandafter\ifx\csname l@#1\endcsname\relax
\typeout{** WARNING: IEEEtran.bst: No hyphenation pattern has been}%
\typeout{** loaded for the language `#1'. Using the pattern for}%
\typeout{** the default language instead.}%
\else
\language=\csname l@#1\endcsname
\fi
#2}}
\providecommand{\BIBdecl}{\relax}
\BIBdecl

\bibitem{wilkes2025quantum}
S.~Wilkes, S.~Brawley, S.~Benjamin, S.~Brierley, W.~Chalupczak, D.~Cielecki, T.~Cubitt, M.~Haji, A.~Hammond, J.~Hance \emph{et~al.}, ``Quantum computing, sensing, and communications,'' \emph{POSTnote}, 2025.

\bibitem{tan2024simultaneous}
P.~Tan, T.~Wang, H.~Zhao, Z.~Tan, P.~Huang, and G.~Zeng, ``Simultaneous quantum and classical communication via multiparameter modulation,'' \emph{Physical Review A}, vol. 109, no.~3, p. 032621, 2024.

\bibitem{orieux2016recent}
A.~Orieux and E.~Diamanti, ``Recent advances on integrated quantum communications,'' \emph{Journal of Optics}, vol.~18, no.~8, p. 083002, 2016.

\bibitem{henshaw2017challenges}
M.~J. d.~C. Henshaw, M.~J. Everitt, V.~M. Dwyer, J.~Lemon, and S.~C. Jones, ``The challenges for systems engineers of non-classical quantum technologies,'' \emph{arXiv preprint arXiv:1710.05643}, 2017.

\bibitem{everitt2016quantum}
M.~J. Everitt, J.~D.~C. Michael, and V.~M. Dwyer, ``Quantum systems engineering: A structured approach to accelerating the development of a quantum technology industry,'' in \emph{2016 18th International Conference on Transparent Optical Networks (ICTON)}.\hskip 1em plus 0.5em minus 0.4em\relax IEEE, 2016, pp. 1--4.

\bibitem{IET2024QuantumSE}
\BIBentryALTinterwordspacing
R.~Claridge, R.~Hancock, D.~Harvey, G.~Marshall, T.~Rabbets, J.~Vovrosh, and P.~Yates. (2024) Quantum technologies: a new frontier for systems engineering? The Institution of Engineering and Technology (IET). Accessed 16 June 2025. [Online]. Available: \url{https://www.theiet.org/impact-society/policy-and-public-affairs/digital-futures-policy/reports-and-papers/quantum-technologies-a-new-frontier-for-systems-engineering}
\BIBentrySTDinterwordspacing

\bibitem{chen2009field}
W.~Chen, Z.-F. Han, T.~Zhang, H.~Wen, Z.-Q. Yin, F.-X. Xu, Q.-L. Wu, Y.~Liu, Y.~Zhang, X.-F. Mo \emph{et~al.}, ``Field experiment on a “star type” metropolitan quantum key distribution network,'' \emph{IEEE Photonics Technology Letters}, vol.~21, no.~9, pp. 575--577, 2009.

\bibitem{peev2009secoqc}
M.~Peev, C.~Pacher, R.~All{\'e}aume, C.~Barreiro, J.~Bouda, W.~Boxleitner, T.~Debuisschert, E.~Diamanti, M.~Dianati, J.~F. Dynes \emph{et~al.}, ``The secoqc quantum key distribution network in vienna,'' \emph{New journal of physics}, vol.~11, no.~7, p. 075001, 2009.

\bibitem{sasaki2011field}
M.~Sasaki, M.~Fujiwara, H.~Ishizuka, W.~Klaus, K.~Wakui, M.~Takeoka, S.~Miki, T.~Yamashita, Z.~Wang, A.~Tanaka \emph{et~al.}, ``Field test of quantum key distribution in the tokyo qkd network,'' \emph{Optics express}, vol.~19, no.~11, pp. 10\,387--10\,409, 2011.

\bibitem{huang2025fully}
C.~Huang, R.~Guan, X.~Liu, S.~Li, W.~He, H.~Liang, Z.~Luo, Z.~Zhang, W.~Li, and K.~Wei, ``Fully connected twin-field quantum key distribution network,'' \emph{arXiv e-prints}, pp. arXiv--2504, 2025.

\bibitem{dianati2008architecture}
M.~Dianati, R.~All{\'e}aume, M.~Gagnaire, and X.~Shen, ``Architecture and protocols of the future european quantum key distribution network,'' \emph{Security and Communication Networks}, vol.~1, no.~1, pp. 57--74, 2008.

\bibitem{BennettBrassard1984}
C.~H. Bennett and G.~Brassard, ``Quantum cryptography: Public key distribution and coin tossing,'' in \emph{Proceedings of the IEEE International Conference on Computers, Systems \& Signal Processing}.\hskip 1em plus 0.5em minus 0.4em\relax Bangalore, India: IEEE, 1984, pp. 175--179.

\bibitem{ekert1991quantum}
A.~K. Ekert, ``Quantum cryptography based on bell’s theorem,'' \emph{Physical review letters}, vol.~67, no.~6, p. 661, 1991.

\bibitem{Lo2012}
H.~K. Lo, M.~Curty, and B.~Qi, ``Measurement-device-independent quantum key distribution,'' \emph{Physical Review Letters}, vol. 108, 3 2012.

\bibitem{iso15288:2023}
\emph{{ISO/IEC/IEEE 15288:2023 — Systems and software engineering — System life cycle processes}}, \url{https://www.iso.org/standard/78832.html}, ISO/IEC/IEEE Std., 2023, standard published jointly by ISO, IEC and IEEE.

\bibitem{hause2019model}
M.~Hause and J.~Hummell, ``Model-based product line engineering--enabling product families with variants,'' \emph{INSIGHT}, vol.~22, no.~2, pp. 43--48, 2019.

\bibitem{li_model-based_2015}
M.~Li, F.~Batmaz, L.~Guan, A.~Grigg, M.~Ingham, and P.~Bull, ``Model-based systems engineering with requirements variability for embedded real-time systems,'' in \emph{2015 IEEE International Model-Driven Requirements Engineering Workshop (MoDRE)}.\hskip 1em plus 0.5em minus 0.4em\relax IEEE, 2015, pp. 1--10.

\bibitem{schafer_variability_2021}
A.~Sch{\"a}fer, M.~Becker, M.~Andres, T.~Kistenfeger, and F.~Rohlf, ``Variability realization in model-based system engineering using software product line techniques: an industrial perspective,'' in \emph{Proceedings of the 25th ACM International Systems and Software Product Line Conference-Volume A}, 2021, pp. 25--34.

\bibitem{moiseev2025optical}
S.~A. Moiseev, M.~M. Minnegaliev, K.~I. Gerasimov, E.~S. Moiseev, A.~D. Deev, and Y.~Y. Balega, ``Optical quantum memory in atomic ensembles: physical principles, experiments, and potential of application in a quantum repeater,'' \emph{Uspekhi Fizicheskikh Nauk}, vol. 195, no.~5, pp. 455--477, 2025.

\bibitem{granelli2022novel}
F.~Granelli, R.~Bassoli, J.~N{\"o}tzel, F.~H. Fitzek, H.~Boche, and N.~L. da~Fonseca, ``A novel architecture for future classical-quantum communication networks,'' \emph{Wireless Communications and Mobile Computing}, vol. 2022, no.~1, p. 3770994, 2022.

\bibitem{azuma2023quantum}
K.~Azuma, S.~E. Economou, D.~Elkouss, P.~Hilaire, L.~Jiang, H.-K. Lo, and I.~Tzitrin, ``Quantum repeaters: From quantum networks to the quantum internet,'' \emph{Reviews of Modern Physics}, vol.~95, no.~4, p. 045006, 2023.

\bibitem{liorni2021quantum}
C.~Liorni, H.~Kampermann, and D.~Bru{\ss}, ``Quantum repeaters in space,'' \emph{New Journal of Physics}, vol.~23, no.~5, p. 053021, 2021.

\bibitem{ji2022structure}
S.~Ji, M.~Wilkinson, and C.~E. Dickerson, ``Structure preserving transformations for practical model-based systems engineering,'' in \emph{2022 IEEE International Symposium on Systems Engineering (ISSE)}.\hskip 1em plus 0.5em minus 0.4em\relax IEEE, 2022, pp. 1--8.

\bibitem{ji2023requirements}
S.~Ji, C.~E. Dickerson, and M.~Wilkinson, ``Requirements rationalization and synthesis enabled by model synchronization,'' \emph{IEEE Open Journal of Systems Engineering}, vol.~1, pp. 26--37, 2023.

\bibitem{li2019product}
M.~Li, A.~Grigg, C.~Dickerson, L.~Guan, and S.~Ji, ``A product line systems engineering process for variability identification and reduction,'' \emph{IEEE Systems Journal}, vol.~13, no.~4, pp. 3663--3674, 2019.

\bibitem{bjergstrom2020quantum}
K.~Bjergstrom, ``Quantum systems engineering,'' Ph.D. dissertation, Loughborough University, 2020.

\bibitem{Feynman1982}
R.~P. Feynman, ``Simulating physics with computers,'' Tech. Rep., 1982.

\bibitem{Troyer2005}
M.~Troyer and U.~J. Wiese, ``Computational complexity and fundamental limitations to fermionic quantum monte carlo simulations,'' \emph{Physical Review Letters}, vol.~94, 5 2005.

\bibitem{iso15288:2015}
\emph{{ISO/IEC/IEEE 15288:2015 — Systems and software engineering — System life cycle processes}}, \url{https://www.iso.org/standard/63711.html}, ISO/IEC/IEEE Std., 2015, standard published jointly by ISO, IEC and IEEE.

\bibitem{Silverman2023}
S.~J. Silverman and T.~Jiron, ``Quantum mbse and quantum sysml,'' in \emph{MILCOM 2023 - 2023 IEEE Military Communications Conference: Communications Supporting Military Operations in a Contested Environment}.\hskip 1em plus 0.5em minus 0.4em\relax Institute of Electrical and Electronics Engineers Inc., 2023, pp. 95--99.

\bibitem{perez2022design}
R.~P{\'e}rez-Castillo and M.~Piattini, ``Design of classical-quantum systems with uml,'' \emph{Computing}, vol. 104, no.~11, pp. 2375--2403, 2022.

\bibitem{guo2025quanuml}
X.~Guo, S.~Saito, and J.~Zhao, ``Quanuml: Towards a modeling language for model-driven quantum software development,'' \emph{arXiv preprint arXiv:2506.04639}, 2025.

\bibitem{dumitrescu_capturing_2014}
C.~Dumitrescu, P.~Tessier, C.~Salinesi, S.~Gerard, A.~Dauron, and R.~Mazo, ``Capturing variability in model based systems engineering,'' in \emph{Complex Systems Design \& Management: Proceedings of the Fourth International Conference on Complex Systems Design \& Management CSD\&M 2013}, 2014, pp. 125--139.

\bibitem{Ishida2025QCNModels}
\BIBentryALTinterwordspacing
H.~Ishida and A.~Elsokary, ``Sysml v1 models for quantum‑key‑distribution (qkd) use‑cases,'' Loughborough Quantum Systems Engineering Research Group, 2025, gitHub repository, accessed 13 June 2025. [Online]. Available: \url{https://github.com/Loughborough-Quantum-System-Engineering/QCN-models}
\BIBentrySTDinterwordspacing

\end{thebibliography}
\end{document}